\documentclass[12pt]{iopart}
\usepackage{iopams}  
\usepackage{graphicx}
\usepackage{bm}
\usepackage{wasysym}
\usepackage{ulem}

\usepackage{amssymb}
\usepackage{color}

\begin{document}

\title[Quantum conditional probabilities]{Quantum conditional probabilities}

\author{Ignacio P\'erez and Alfredo Luis}

\address{Departamento de \'{O}ptica, Facultad de Ciencias
F\'{\i}sicas, Universidad Complutense, 28040 Madrid, Spain}
\ead{alluis@fis.ucm.es}
\vspace{10pt}
\begin{indented}
\item[] April 2022
\end{indented}

\begin{abstract}
We investigate the consistency of conditional quantum probabilities. This is whether there is compatibility between the Kolmogorov-Bayes conditional probabilities and the Born rule. We show that they are not compatible in the sense that there are situations where there is no legitimate density matrix that may reproduce the conditional statistics of the other observable via the Born rule. This is to say that the Gleason theorem does not apply to conditional probabilities. Moreover, we show that when this occurs the joint statistics is nonclassical. We show that conditional probabilities are not equivalent to state reduction, so these results do not affect the validity of the L\"{u}ders expression.
\end{abstract}

\vspace{2pc}

%\submitto{\jpa}

\section{Introduction}

Measurement statistics is at the hearth of the very foundations of the quantum theory. This is so that the real signature of most quantum effects manifests itself in the differences between quantum and classical behavior of probabilities. Examples of this are for example Bell tests \cite{LB90,JB64,AF82}, or plain nonclassicality expressed in terms of lack of a well-behaved distribution, as exemplified by the Glauber-Sudarshan distribution in quantum optics \cite{MW95}. In many occasions quantum-classical discrepancies are actually expressed in terms of intended probability distributions taking negative values in the quantum realm \cite{AF82,MW95,RS08,AR15,BKO16,RPF87,RPF82,RHLSJLLLL09,HFH12,SIH16}.  
\bigskip

In this work we present one of these quantum-classical statistical discrepancies in terms of the inconsistency between the standard Kolmogorov-Bayes formulation of statistics and the Born rule. We consider a joint noisy observation of two incompatible observables leading to a true joint probability distribution $p(x,y)$. We focus on the conditional probability $p(x|y) = p(x,y)/p(y)$ for $x$ given $y$. We show situations where there is no legitimate density matrix $\rho_y$ that might reproduce the conditional probability $p(x|y)$ via the Born rule, $p(x|y)= \mathrm{tr} [\rho_y \Delta (x) ]$, being  $\Delta (x)$ the corresponding positive operator-valued measure (POVM) for the $x$ variable. That is to say that quantum conditional probabilities do not follow Gleason’s theorem \cite{AMG57}. Throughout the work we will focus on POVMs because it is the most general scenario possible. In particular, this includes the case of observables well represented by Hermitian operators. Moreover, the POVM framework is the best suited for noisy joint observation of incompatible observables.

\bigskip

Moreover, we show that when there is no legitimate density matrix $\rho_y$ we have that the joint statistics $p(x,y)$ is nonclassical. This is because once we remove the quantum noise in $p(x,y)$ the so obtained noise-free joint distribution $P(x,y)$ takes negative values. By nonclassical we refer to properties that cannot be explained by the classical theory. Precisely, the most powerful and standard way of revealing nonclassical features is by the lack of a true joint distribution for complementary observables, as it may be the case of the Glauber-Sudarshan $P$ distribution in quantum optics \cite{MW95}. Usually this is in the form of observation processes that in classical physics would lead to a legitimate joint probability distribution, but that in quantum physics fail to provide it. This is the idea of the formalism that we will follow in this work, that has been presented and extensively discussed in Refs. \cite{MAL20,GBAL18,AL16a,AL16b,LM17,CL20,MAL21}. It is worth noting also that our work is not about the definition or construction of exact joint probabilities $P(x,y)$, but about conditional probabilities once the noisy joint distribution $p(x,y)$ has been obtained by practical means. The exact joint distribution $P(x,y)$ is just invoked as evidence of nonclassical behavior.

\bigskip

We will show in detail later that this situation is different in general from the problem of quantum state reduction associated with measurement, as clearly illustrated by a trivial example in Sec. II and the practical case of Young interferometer and homodyne detection examined in Appendices A and B. Note that if the solution of the problem would be given by L\"{u}ders approach to state reduction, then $\rho_y$ would be always a well defined density matrix, and we show that this is not always the case. So, this is not a debate  about L\"{u}ders expression since our work would shows that conditional probabilities are not equivalent to state reduction \cite{IG13,BL09,MO85}.  

\bigskip

The existence of joint probabilities is an old and interesting problem as far as probability for two noncommuting observables is actually the next important case in the quantum statistical connection after the probability of a single observable \cite{AF82,JU94,AD18}. Moreover, it has a direct relation with some of the most celebrated theorems at the hearth of the very foundations of the quantum theory, such as the Bell and Kochen-Specker theorems, as well as with potential quantum generalizations of conventional probability theory \cite{AF82a,IP82,SPG84,JDM04} .

\bigskip

Other formulations of the problem of quantum conditional probabilities can be found in Refs. \cite{JDM04,DJM97}, via Kirkwood-type relations. This is to say that if $\Delta_X (x)$ and $\Delta_Y (y)$ are projectors determining the probability that two observables $X$, $Y$ take the values $x,y$, respectively, then a joint distribution for both with correct marginals is constructed via their product in the form $\Delta_{X,Y} (x,y) \propto \Delta_X (x) \Delta_Y (y)$, which in general is neither Hermitian nor positive semidefinite \cite{JGK33,PAMD45,HWL95}. Nevertheless, Kirkwood-type expressions have demonstrated very interesting properties in quantum nonclassical statistics \cite{HFH12}. Other proposal within the context of quantum information focus on the relation between completely positive maps and compound states \cite{AKMS06,MSL06}. A further different notion of quantum conditional probability has been addressed in  \cite{CA99}, based on the conditional von Neumann entropy, while in  \cite{BK21} a novel form of quantum conditional probability is introduced to define new measures of quantum information in a dynamical context, revealing a failure of the Bayes theorem.

\bigskip

 In Sec. II we provide the main settings and definitions required. Then, for the sake of simplicity, throughout the work we focus on a qubit system. In Sec. III we address the solution to the problem posed, showing that for every POVM there are states such that the Kolmogorov-Bayes conditional probabilities and the Born rule are incompatible. Then in Sec. IV we relate these facts with the quantum nature of the corresponding joint statistics. Finally, in the Appendices A and B we provide a real physical implementation of the analysis carried out in the rest of the paper in terms of fully measurable quantities. In Appendix A this is a Young interferometer, where the abstract observables $X$, $Y$ become the classic complementary observables of interferometric path and interference. This provides physical intuition in terms of the usual quantum linear optics with a photon passing through a Young interferometer, phase plates and linear polarizers. In the Appendix B we consider the case of double homodyne detection implemented via beam splitting and interference. Moreover, the Appendices also serve to show that conventional state reduction in a Neumark scenario \cite{CWH76,MO97,AP93}, also spelled Naimark in the literature, does not solve the question addressed here. We can recall that every POVM can be represented by a family of orthogonal projectors on an enlarged Hilbert space, that includes ancillary degrees of freedom where partial information about the observed system is transferred in the measurement process. Beyond qubit systems, in Appendix B we provide the example of double homodyne detection in the infinite-dimensional space of a single-mode field, proving the validity of the main results of this work.

\bigskip

\section{Settings}

Let us consider the joint measurement of two observables, which we call $X$ and $Y$, of a given physical system, $S$. The measurement should admit a joint probability distribution $p(x,y)$, that tells us the probability that the property $X$ takes the value $x$ and property $Y$ takes the value $y$. This is to say that $X$ and $Y$ must be compatible in the sense of being jointly measurable \cite{PM68}, also referred to as commeasurable \cite{POVM} or coexistent \cite{PB86}, which in general is not equivalent to commuting. Throughout this work we consider that $X$ and $Y$ actually provide the noisy joint observation of two incompatible observables, say the Pauli matrices $\sigma_X$ and $\sigma_Y$ for a qubit system.

\bigskip

According to the Born rule and the Gleason theorem, every $p(x,y)$ is linearly determined from the density-matrix of the system $\rho$ via a positive operator-valued measure $\Delta (x,y)$ as 
\begin{equation}
\label{js}
    p (x,y) = \mathrm{tr} \left [\rho \Delta (x,y) \right ] ,
\end{equation}
where the statistical interpretation demands that 
\begin{equation}
\label{povm}
  \Delta^\dagger (x,y) = \Delta (x,y) , \quad  \Delta (x,y) \geq 0, \quad \sum_{x,y} \Delta (x,y) = I, 
\end{equation}
and $I$ is the identity. We will assume a discrete character for $x,y$ just for the sake of simplicity. The corresponding marginal POVMs for the observables $X$ and $Y$ are, respectively,
\begin{equation}
\label{marg}
 \Delta_X (x) = \sum_y \Delta (x,y) , \quad \Delta_Y (y) = \sum_x \Delta (x,y) ,
\end{equation}
so that the marginal probabilities $p_W (w)$ that each observable $W=X,Y$ takes the value $w=x,y$  are 
\begin{equation}
\label{mar}
 p_W (w) = \mathrm{tr} \left [\rho \Delta_W (w) \right ].
\end{equation}

\bigskip

A Kolmogorovian-Bayesian formulation of plain statistics lead us to define the conditional probability distribution $p(x|y)$ of getting $x$ provided that $Y$ takes the value $y$ as
\begin{equation}
\label{Bp}
    p(x,y) = p(x|y)p_Y (y) .
\end{equation}
The main questions we address in this work is whether this conditional probability determines a legitimate and unique density matrix $\rho_y$ such that the conditional probability $p(x|y)$ may be obtained via the Born rule, that is
\begin{equation}
\label{rs0}
 p(x|y) =\mathrm{tr} \left [\rho_y \Delta_X (x) \right ] .
\end{equation}
No further assumptions are made beyond the basic relations (\ref{js}), (\ref{mar}), (\ref{Bp}) and (\ref{rs0}). We even avoid to refer to $\rho_y$ as the reduced state associated with the measurement of $Y$, nor impose any form for the relation between $\rho_y$ and $\rho$, such as complete positivity of  $\rho \rightarrow \rho_y$ mapping, or any other physical requirement. This is because we let the formalism to answer these points by itself once the solution is found. In particular, note that  (\ref{Bp}) is independent of whether the measurement of $X$ precedes the measurement of $Y$ or the other way round, so there is no point here regarding causal ordering. This will be clearly illustrated by the two examples presented in the Appendices.

\bigskip

As a suitable point of comparison let us consider the straightforward case when the joint POVM factorizes as the product of independent POVMs defined in different Hilbert spaces. 

\bigskip

\noindent {\it Theorem 1.--} If the POVM factorizes, this is
\begin{equation}
     \Delta (x,y) = \Delta_X (x) \otimes \Delta_Y (y),
\end{equation}
then we have the following form for $\rho_y$:
\begin{equation}
\label{ss}
    \rho_y = \frac{\mathrm{tr}_Y \left [\rho \Delta_Y (y) \right ]}{\mathrm{tr}_{X \otimes Y} \left [\rho \Delta_Y (y) \right ]} ,
\end{equation}
where for clarity the subscripts on the traces indicate the space where the trace is performed.

\bigskip

\noindent {\it Proof.--} This is demonstrated by a simple ordering of the trace in  (\ref{js}). It can be easily seen that $\rho_y$ is a legitimate density matrix as it can be easily checked, including positive semidifiniteness, that we will represent as $\rho_y \succeq 0$. This is because for any vector $|\psi \rangle$ in the $X$ space we have   
\begin{equation}
    \langle \psi |\rho_y |\psi \rangle \propto \mathrm{tr}_{X \otimes Y} \left [\rho |\psi \rangle\langle \psi |\otimes  \Delta_Y (y) \right ] \geq 0,
\end{equation}
since $\rho \succeq 0$,  $|\psi \rangle\langle \psi | \succeq 0$ and $\Delta_Y (y) \succeq 0$ $\blacksquare$.

\bigskip

This solution in the trivial case points out that we are not in a simple scenario of standard state reduction. In particular, notice that the state $\rho_y$ belongs to the Hilbert space of the $X$ measurement instead of the Hilbert space of the measurement $Y$.

\bigskip

\subsection{Qubit system}

Next we particularize this POVM framework to a qubit system. The chain of though is the following. We first construct the most general POVM $\Delta (x,y)$ that represents a noisy joint observation of the two incompatible observables $\sigma_X$ and $\sigma_Y$. Once we have $\Delta (x,y)$ we derive all the probabilities involved in (\ref{Bp}) for the most general state of the system. Then, in Sec. 3 we will solve Eq. (\ref{Bp}) to obtain $\rho_y$. 

The most simple form for such $\Delta (x,y)$  involves just two dichotomic variables $x,y$, that may be referred to  as the compatible $X,Y$ observables taking the values $x,y$, respectively. By dichotomic we mean that $x,y = \pm 1$. In these conditions, the most general form for $\Delta (x,y)$ is
\begin{equation}
    \Delta (x,y) = \frac{1}{4} \left ( \Delta_0 + x \Delta_X + y \Delta_Y + x y \Delta_{XY} \right ) , 
\end{equation}
where the operators $\Delta_{X,Y,XY}$ no longer depend on $x,y$, being linear functions of the  Pauli matrices $\bsigma$ and the $2 \times 2$ identity matrix $\sigma_0$. From the third condition in  (\ref{povm}) we get $\Delta_0 = \sigma_0$. Moreover, we just add a unbiasedness requirement on $\Delta(x,y)$ \cite{YLLO10}, that is that the maximally mixed state carries no information, wich means that its statistics $p(x,y)$ does not depend on $x,y$:
\begin{equation}
\label{ntc}
    \rho= \frac{1}{2} \sigma_0 \rightarrow p(x,y) = \frac{1}{4} .
\end{equation}
This means that the $\Delta_{X,Y,XY}$  have no $\sigma_0$ component, so that the final form for our POVM is
\begin{equation}
\label{D}
\Delta (x,y) = \frac{1}{4} \left  [ \sigma_0 +  \bi{S} (x,y) \cdot  \bsigma  \right ] ,
\end{equation}
with
\begin{equation}
\label{tS}
\bi{S} (x,y) = x \gamma_X \bi{S}_X + y \gamma_Y \bi{S}_Y+ x y \gamma_{XY} \bi{S}_{XY},  
\end{equation}
where $\bi{S}_X$, $\bi{S}_Y$, and $\bi{S}_{XY}$ are three-dimensional unit-modulus real vectors. The $\gamma$ factors are real and positive expressing the accuracy of the observation as explained below. In any case all these elements are constrained so that  $|\bi{S} (x,y) | \leq 1$, so that conditions (\ref{povm}) are satisfied. For definiteness let us consider that $\bi{S}_X, \bi{S}_Y$ and $\bi{S}_{XY}$ are mutually orthogonal  \begin{equation}
\label{ttr}
    \bi{S}_X = (1,0,0), \quad  \bi{S}_Y = (0,1,0), \quad  \bi{S}_{XY} = (0,0,1), \quad
\end{equation}
so that $|\bi{S} (x,y) | \leq 1$ holds provided that  
\begin{equation}
\gamma_X^2 + \gamma_Y^2 + \gamma_{XY}^2 \leq 1.
\end{equation}
This choice is motivated by our understanding of $\Delta (x,y)$ as a noisy joint measurement of $\sigma_X$ and $\sigma_Y$. In the Appendix A we provide a detailed derivation of this POVM in the particular but significant enough case of path-interference complementarity in a Young interferometer. Incidentally, note that the choice  $\Delta_{XY} =\sigma_Z$ is rather natural if we consider that all information about the observable $\sigma_W$ is in $\Delta_W$ for both $W=X,Y$. Moreover, since  $\Delta_{XY}$ represents correlations between $X$ and $Y$ we can take into account that the most similar relation in terms of $\sigma_X$ and $\sigma_Y$ says that $\sigma_X \sigma_Y = i\sigma_Z$.

\bigskip

The joint POVM $\Delta (x,y)$ leads to the marginals for $X$ and $Y$ 
\begin{equation}
\label{DW}
    \Delta_W (w) = \frac{1}{2} \left  ( 1 +  w \gamma_W \sigma_W \right ) ,
\end{equation}
with $W=X,Y$, $w=x,y$.

\bigskip

The system state can be expressed as 
\begin{equation}
\label{rho}
\rho =\frac{1}{2} \left ( \sigma_0 +\bi{s}\cdot \bsigma \right ) ,
\end{equation}
where
\begin{equation}
    \bi{s} = \mathrm{tr} (\rho \bsigma ) = (s_X,s_Y,s_Z) ,
\end{equation}
is a three-dimensional real vector with  $|\bi{s}| \leq 1$. Then we have the following joint distribution 
\begin{equation}
\label{pxy}
    p (x,y) = \frac{1}{4} \left ( 1+ x \gamma_X s_X +  y \gamma_Y s_Y + x y \gamma_{XY
    } s_Z \right ) ,
\end{equation}
with marginals 
\begin{equation}
\label{pm}
p_X (x) = \frac{1}{2} \left  ( 1 +  x \gamma_X s_X \right ) , \quad 
p_Y (y) = \frac{1}{2} \left  ( 1 +  y \gamma_Y s_ Y \right ) .
\end{equation}

\bigskip

\section{Solution}
Combining  (\ref{Bp}),  (\ref{pxy}) and (\ref{pm}) we readily get 
\begin{equation}
\label{cond1}
p (x|y) = \frac{1}{2} \left  ( 1 +  x \gamma_X \frac{ s_X + y s_Z \gamma_{XY}/\gamma_X}{1 +  y \gamma_Y s_Y} \right ) .
\end{equation}
Expressing the the desired density matrix $\rho_y$ in the most general form
\begin{equation}
\label{rs}
\rho_y =\frac{1}{2} \left ( \sigma_0 +\bi{t}\cdot \bsigma \right ) ,
\end{equation}
and using  (\ref{DW}) we get 
\begin{equation}
\label{cond2}
 p(x|y) =\mathrm{tr} \left [\rho_y \Delta_X (x) \right ] = \frac{1}{2} \left  ( 1 +  x  \gamma_X t_X \right ) .
\end{equation}
Then, equating (\ref{cond1}) and (\ref{cond2}) we obtain
\begin{equation}
\label{trs}
t_X = \frac{ s_X + y s_Z \gamma_{XY}/\gamma_X}{1 +  y \gamma_Y s_Y} .
\end{equation}
This is the only condition on $\bi{t}$ that we can get, so the other components $t_Y$ and $t_Z$ are free parameters, always respecting the condition $|\bi{t}| \leq 1$ for $\rho_y$ being positive semidifinite.

\bigskip

Common quantum intuition may suggest that  $\rho_y$ is the reduced state associated with the measurement of $Y$, typically as $\sqrt{\Delta_Y (y)} \rho \sqrt{\Delta_Y (y)}$ according to the L\"{u}ders rule, taking always into account that the square root of a matrix is not unique reflecting the fact that POVM alone does not fully determine the reduced state . But in our case this is enough to show that there is no simple relation between $\rho$ and $\rho_y$ implemented by a quantum instrument in the typical form of a completely positive linear map. In particular, $\Delta_Y (y)$ does not depend on $\gamma_{XY}$ while $\rho_y$ actually does. Also, the factorized case already analyzed around (\ref{ss}) shows that such intuition fails regarding the particular problem we are addressing here.

\bigskip

The most important point is that there are situations in which there is no legitimate state $\rho_y$. This occurs when (\ref{trs}) is incompatible with the condition $|\bi{t}| \leq 1$ required so that $\rho_y \succeq 0$. We will refer to the case when $\rho_y$ is not positive semidifinite as $\rho_y \nsucceq 0$. For example this is the case of $s_X=s_Y=0$ provided that 
\begin{equation}
\left | t_X \right |= \left | y s_Z \frac{\gamma_{XY}}{\gamma_X} \right | >1 , 
\end{equation}
that holds provided that $\gamma_{XY}\neq 0$ and $\gamma_X$ is small enough, a choice that is always possible. For details regarding these conditions from the point of view of measurable quantities in a real physical scenario we refer to the Appendix A. Finally, let us prove the following theorem: 

\bigskip

\noindent {\it Theorem 2.--} For all POVMs (\ref{D}) with the choice (\ref{tS}) there is an state $\rho$ such that the corresponding $\rho_y$ is not a legitimate density matrix since $\rho_y \nsucceq 0$.

\bigskip

\noindent {\it Proof.--} Let the state $\rho$ be of the form in  (\ref{rho}) with 
\begin{equation}
    \bi{s} = \left ( \sqrt{1-\gamma_Y^2}, \gamma_Y,0 \right ) ,
\end{equation}
and let it be $y=-1$. Then after  (\ref{trs})
\begin{equation}
    t_X = \frac{1}{\sqrt{1-\gamma_Y^2}} > 1 \rightarrow  \rho_y \nsucceq 0.   \quad \blacksquare
\end{equation}

\bigskip

\section{Nonexistence and nonclassicality}

Let us show here that the nonexistence of $\rho_y$ in  (\ref{trs}) implies the nonclassical character of $\rho$ as next specified briefly recalling the formalism already presented in  \cite{MAL20,GBAL18,AL16a,AL16b,LM17,CL20,MAL21}.

\subsection{Nonclassicalliy}

In a classical-physics scenario, the joint statistics of two observables is of the form, as extensively discussed when looking for test for quantum nonlocality and noncontextuality \cite{MA84,AK00,HP04,AM08,TN11,AK14,WMM14},
\begin{equation}
\label{sep}
    p(x,y)= \sum_{\lambda} \nu_X (x |\lambda) \nu_Y (y |\lambda ) P (\lambda) ,
\end{equation}
where $P (\lambda)$ is a well-behaved probability distribution over the states of the system, denoted by $\lambda$, which typically corresponds to its phase space, and $\nu_X (x |\lambda)$, $ \nu_Y (y, |\lambda )$ are suitable conditional probability distributions defined by the observation process, so they can be known in practice independently of the system state, in our case they will be determined later in  (\ref{fac1}). We say that $ p(x,y)$ is separable when it admits the form (\ref{sep}) for a suitable probability distribution $P (\lambda)$, and two conditional probabilities $\nu_X (x |\lambda)$, $ \nu_Y (y, |\lambda )$, consistent with the measurement at hand. Otherwise, we say that $ p(x,y)$ is nonseparable. It is relevant to stress that this separability condition holds in classical physics respectively of whether the observation is exact or noisy. But in quantum physics the situation is utterly different \cite{MAL21}. Thus we may say that if $p(x,y)$ fails to be separable then it is nonclassical. 

\bigskip

The idea is that the POVM $\Delta (x,y)$ in  (\ref{D}) provides a simultaneous noisy observation of the two incompatible observables $\sigma_X$ and $\sigma_Y$. Because of this in Eq. (\ref{sep}) we are not addressing a full hidden-variable theory valid for arbitrary measuring settings, but just a classical-like model for the particular noisy measuring procedure leading to the POVM (\ref{D}). In this spirit we consider the four combinations $\lambda = (x,y)$ with $x,y = \pm 1$ to represent the exact values of the observables $\sigma_X$ and $\sigma_Y$, so that $P(x,y)$ in Eq. (\ref{sep}) represents the exact noiseless joint distribution for these observables. In this spirit we may also naturally assume that the conditional probabilities take the form $\nu_X (x|\lambda)= \nu_X (x|x^\prime)$  and $\nu_Y (y|\lambda)= \nu_y(y|y^\prime)$, so that the statistics of $X$ is derived exclusively in terms of the statistics of $\sigma_X$, and equivalently, the statistics of $Y$ is derived exclusively in terms of the statistics of $\sigma_Y$. Moreover, we assume that the observed marginals $p_X (x)$ and  $p_Y (y)$ in  (\ref{pm}) contain complete statistical information about $\sigma_X$ and $\sigma_Y$ in the state $\rho$. To provide suitable expressions for the conditional probabilities, $\nu_X (x|x^\prime)$  and $\nu_y(y|y^\prime)$ we focus on the observed marginal statistics in  (\ref{pm}) considering the case of eigenstates of $\sigma_X$ and $\sigma_Y$, so we may conclude that 
\begin{equation}
\label{fac1}
   \nu_W (w|w^\prime )= \frac{1}{2} \left ( 1+ \gamma_W w w^\prime \right ), 
\end{equation}
for $W=X,Y$ and $w= x,y$, $w^\prime = x^\prime,y^\prime$. Therefore, in a classical scenario the observed joint statistics satisfies a separability condition of the form 
\begin{equation}
\label{dir}
    p(x,y)= \sum_{x^\prime, y^\prime = \pm 1} \nu_X (x | x^\prime) \nu_Y (y | y^\prime) P (x^\prime, y^\prime) .
\end{equation}

\bigskip

A direct way to test separability is to invert relation (\ref{dir}) to obtain $P(x,y)$ in terms of the observed statistics $p(x,y)$: if $P(x,y) \ngeq 0$ then $p(x,y)$ is not separable. This can be done in the form 
\begin{equation}
\label{inv}
    P(x,y)= \sum_{x^\prime, y^\prime = \pm 1} \mu_X (x, x^\prime) \mu_Y (y, y^\prime) p (x^\prime, y^\prime) ,
\end{equation}
where $\mu_W (w, w^\prime)$ are the inverses of $\nu_W (w| w^\prime)$
\begin{equation}
\label{muW}
\mu_W (w,w^\prime) = \frac{1}{2} \left ( 1 + \frac{w w^\prime}{ \gamma_W}  \right ), 
\end{equation}
leading to 
\begin{equation}
\label{Pxy}
    P (x,y) = \frac{1}{4 } \left ( 1+ x s_X +  y s_Y + x y s_Z \frac{\gamma_{XY}}{\gamma_X \gamma_Y} \right ) .
\end{equation}
Thus we may refer to the case $ P(x,y) \ngeq 0$ as evidence of nonclassical behavior. For a parallel inversion analysis in an slightly different context see for example  \cite{SIH16}.

\bigskip

\subsection{Lack of density matrix $\rho_y$ implies nonclassicallity}

\noindent {\it Theorem 3.--} If $\rho_y$ is not positive semidefinite then  $ P(x,y)$ cannot be a probability distribution, this is $\rho_y \nsucceq 0 \rightarrow P(x,y) \ngeq 0$.

\bigskip

\noindent {\it Proof.--} 
Let us begin with considering $t_x>0$ so that $\rho_y \nsucceq 0$ holds when $t_x>1$. After  (\ref{trs}) we have that $t_x>1$ implies 
\begin{equation}
    1-s_X + y\gamma_Y s_Y - y s_Z \frac{\gamma_{XY}}{\gamma_X }  <0 .
\end{equation}
Since always $1-s_X \geq 0$ then
\begin{equation}
    y \gamma_Y s_Y - y s_Z \frac{\gamma_{XY}}{\gamma_X }  <0 ,
\end{equation}
and  then, since $1 \geq \gamma_Y >0 $
\begin{equation}
y  s_Y - y s_Z \frac{\gamma_{XY}}{\gamma_X \gamma_Y}  \leq y \gamma_Y  s_Y - y s_Z \frac{\gamma_{XY}}{\gamma_X} ,
\end{equation}
which implies
\begin{equation}
\label{ins}
    1-s_X + y s_Y - s_Z \frac{\gamma_{XY}}{\gamma_X \gamma_Y} <0 .
\end{equation}
This readily means after  (\ref{Pxy}) that  $P(x=-1,y)<0$. Likewise  $t_X < -1$ implies that $P(x=1,y)<0$. $\blacksquare$

\bigskip

This is to say that the incompatibility of the Kolmogorov-Bayes conditional probability (\ref{Bp}) and the Born rule is a nonclassical phenomenon. This might be a signature of the intrinsic nonclassical nature of the measurement processes, as already pointed out in Ref.\cite{LA20}, implying a failure of the classical-like models such as (\ref{sep}).

\bigskip

The other way round there are situations with $P(x,y) \ngeq 0$ but $|t_x|\leq 1$. For example this is the case of $s_X=s_Y=0$ provided that 
\begin{equation}
\frac{1}{\gamma_Y} \geq \left | s_Z \frac{\gamma_{XY}}{\gamma_X \gamma_Y} \right | >1 .
\end{equation}

\bigskip

We may nevertheless derive a complete equivalence between $\rho_y \nsucceq 0$ and $P(x,y) \ngeq 0$ provided we examine the same problem of compatibility between conditional probabilities and Born rule after data inversion. This is replacing  (\ref{Bp}) by 
\begin{equation}
\label{BP}
    P(x,y) = P(x|y)P_Y (y) , 
\end{equation}
where 
\begin{equation}
 P_Y(y) =\sum_x P(x,y) = \mathrm{tr} \left [\rho \Omega_Y (y) \right ] ,
\end{equation}
and
\begin{equation}
 \Omega_W (w) =  \frac{1}{2} \left ( 1+ w \sigma_W \right ) ,
\end{equation}
are the actual orthogonal projectors on the eigenstates of $\sigma_W$, $W=X,Y$, $w=x,y$ so that 
\begin{equation}
\label{PWw}
P_W (w) = \frac{1}{2} \left  ( 1 +  w s_W \right )  .
\end{equation}
Then we look for the density matrix $\rho_y$ defined by the Gleason-type relation
\begin{equation}
\label{irhoy}
 P(x|y) =\mathrm{tr} \left [\rho_y \Omega_X (x) \right ] .
\end{equation}

\bigskip

\noindent {\it Theorem 4.--} There is a complete equivalence between $\rho_y$ not being positive semidefinite and $P(x,y)$ not being a probability distribution, this is $\rho_y \nsucceq 0$ if and only if $P(x,y) \ngeq 0$. 

\bigskip{
\noindent }{\it Proof.--} After  (\ref{Pxy}) and (\ref{PWw}) we get 
\begin{equation}
P (x|y) = \frac{1}{2} \left  ( 1 +  x \frac{ s_X + y s_Z \frac{\gamma_{XY}}{\gamma_X \gamma_Y}}{1 +  y s_Y} \right ) 
\end{equation}
so that asking for the $\rho_y$ density matrix in  (\ref{irhoy}) using the same notation in (\ref{rs}), we get 
\begin{equation}
t_X = \frac{ s_X + y s_Z \frac{\gamma_{XY}}{\gamma_X \gamma_Y}}{1 +  y  s_Y} ,
\end{equation}
so that in this case we readily get that  $|t_X| > 1$ holds if and if and only if $P(x=1,y)<0$ or $P(x=-1,y)<0$, so there is no legitimate $\rho_y$ if and only if the joint statistics is nonclassical  $\blacksquare$. 

Roughly speaking the equivalence is clear after  (\ref{BP}) since $\Omega (x) \succeq 0$ and $P_Y(y) \geq 0$, so that the negativity of the inverted distribution $P(x,y) \ngeq 0$ must come from that $\rho_y \nsucceq 0$.

\bigskip

It might seem strange to invoke negative distributions in this statistical context, but we might recall that there are many formalisms where the peculiarities of quantum mechanics are expressed in this way, or even in terms of complex distributions \cite{AF82,MW95,RS08,AR15,BKO16,RPF87,RPF82,RHLSJLLLL09,SIH16}, in particular  regarding conditional probabilities \cite{HFH12}.

\bigskip

\section{Conclusions}

We have examined the compatibility between classical-like Kolmogorov-Bayes conditional probabilities, Born rule and Gleason theorem for POVMs representing  noisy joint observations of incompatible observables. We have shown that there are situations where they are incompatible. We have found that this incompatibility has a clear and definite relation with the nonclassicality of the statistics defined by the very same POVM, manifesting in a nonseparable joint statsitics. 

\bigskip

\section*{Acknowledgments}
We acknowledge financial support from Spanish Ministerio de Econom\'ia y Competitividad Project No. FIS2016-75199-P. We thank Dr. G. Garc\'ia Moreno for continuous support and helpful comments.

\appendix

\section{Young interferometer realization}
Path-interference duality is a classic example of complementarity at work in a two-dimensional space, so it is a natural arena for a nontrivial joint observation of the kind we are addressing in this work. To this end we will follow the scheme in   \cite{GBAL18}. To fix ideas let us assume that we are dealing with a single-photon state of light. 

\bigskip

Let us represent the state at the apertures of a Young interferometer by the two orthogonal kets $|\pm \rangle$ that represent the presence of the photon at the corresponding aperture. They may be always regarded as the eigenvectors of the Pauli matrix $\sigma_Y$, so let it be $Y \propto \sigma_Y$ as the path observable. 

\bigskip

Interference occurs by the coherent superposition of $|\pm \rangle$, so we may represent interference by the observable $X \propto\sigma_X$. We will consider that the observed interference $X$ is directly measured on the system space $S$ by projection on the eigenstates $|x \rangle$ of $\sigma_X$. 

\bigskip

Path information, this is information about $Y$,  will be transferred from the system space to an auxiliary space $A$, that will be the polarization of the photon at each aperture. The path information can be imprinted in the polarization state by a different phase plate placed on each aperture. When the apertures are illuminated by right-handed circularly polarized light, represented by the vector $|\circlearrowright \rangle$ in the polarization space $A$, the phase plates produce the following aperture-dependent polarization transformation  
\begin{equation}
\label{tfmt}
    |\pm\rangle \otimes |\circlearrowright \rangle \rightarrow |\pm\rangle \otimes|\pm \theta \rangle ,
\end{equation}
being 
\begin{equation}
    |\pm \theta \rangle = \cos \frac{\theta}{2} |\circlearrowright \rangle \pm \sin \frac{\theta}{2} |\circlearrowleft \rangle .
\end{equation}
This corresponds to the action of the unitary operator 
\begin{equation}
     U|\pm\rangle \otimes |\circlearrowright \rangle = |\pm\rangle \otimes |\pm \theta \rangle ,
\end{equation}
with
\begin{equation}
     U= \rme^{-\rmi\frac{\theta}{2}\sigma_Y \otimes \Sigma_2}  = \cos \frac{\theta}{2} \sigma_0\otimes\Sigma_0 - \rmi \sin \frac{\theta}{2} \sigma_Y \otimes \Sigma_2 ,
\end{equation}
where for the sake of clarity we denote by capital $\Sigma$ the Pauli matrices in the auxiliary polarization space $A$, being $|\circlearrowright \rangle $ and $ |\circlearrowleft \rangle $ the eigenvectors of $\Sigma_3$, with eigenvalues $+1$ and $-1$ respectively.

\bigskip

Then, the path information is retrieved by measuring any combination of the  observables represented by the Pauli matrices $\Sigma_1$ and $\Sigma_2$ in the auxiliary polarization space, say
\begin{equation}
    \Sigma_\varphi = \cos \varphi \Sigma_1 + \sin \varphi \Sigma_2 .
\end{equation}
This polarization measurement can be very easily achieved in practice with a linear polarizer, where $\varphi$ represents the orientation of its axis. We denote as $|y \rangle$ the eigenvectors of  $\Sigma_\varphi$ with eigenvalue $y=\pm 1$, so the photon passing through the polarizer is represented by the vector $|y = 1 \rangle$ while the photon being stopped by the polarizer is represented by the vector $|y=-1 \rangle$.
 
\bigskip
Let us denote by $\tilde{\Delta} (x,y)$ the actual projection-valued measure in the total space $S \otimes A$
\begin{equation}
    \tilde{\Delta} (x,y)= |x\rangle \langle x| \otimes |y\rangle \langle y|,
\end{equation}
leading to the joint statistics 
\begin{equation}
\label{sf}
    p(x,y) = \mathrm{tr}_{S \otimes A}\left [  U \rho \otimes \rho_A U^\dagger \tilde{\Delta} (x,y) \right ] =
     \mathrm{tr}_S \left [  \rho \Delta (x,y) \right ] ,
\end{equation}
where 
\begin{equation}
\rho_A = |\circlearrowright \rangle \langle \circlearrowright |,
\end{equation}
and the subscripts $A$ and $S$ refers in each case to the auxiliary polarization space or the system space, and we use no subscript $S$ in $\rho$ and $\Delta (x,y)$ to match the notation in the main body of the text. Then the actual POVM  $\Delta (x,y)$ just defined in the system space $S$ is 
\begin{equation}
\Delta (x,y) = \mathrm{tr}_A \left [   \rho_A U^\dagger \tilde{\Delta} (x,y) U \right ] ,
\end{equation}
which exactly of the form in (\ref{D}) and (\ref{tS}) with 
\begin{equation}
    \gamma_X = \cos \theta ,\quad \gamma_Y =  \sin \theta \cos \varphi, \quad \gamma_{XY}=  \sin \theta \sin \varphi,
\end{equation}
that are actually points on the surface of a unit sphere, 
\begin{equation}
    \gamma_X^2+\gamma_Y^2+\gamma_{XY}^2 = 1 .
\end{equation}

\bigskip

Let us investigate the reduced states $\tilde{\rho}_y$ associated with the $Y$ measurement, with a significant proviso: This is not what we were looking for in the main body of the paper, i. e. $\rho_y \neq \tilde{\rho}_y$, mainly because 
\begin{equation}
  \Delta_W (w) \neq |w \rangle \langle w |,
\end{equation}
for $W=X,Y$, $w=x,y$. To determine $\tilde{\rho}_y$ we follow exactly the same steps mentioned in the case of a factorized POVM since $X$ and $Y$ have their own independent Hilbert spaces,  
\begin{equation}
    \tilde{\rho}_y= \frac{\mathrm{tr}_A \left [U \rho \otimes \rho_A U^\dagger | y \rangle \langle y | \right ]}{\mathrm{tr}_{S \otimes A} \left [U \rho \otimes \rho_A U^\dagger | y \rangle \langle y | \right ]} ,
\end{equation}
leading to 
\begin{equation}
     \tilde{\rho}_y = \frac{\Lambda_y \rho \Lambda^\dagger_y}{\mathrm{tr}_S \left [\Lambda^\dagger_y \Lambda_y \rho \right ]} ,
\end{equation}
with 
\begin{equation}
    \Lambda_y = \langle y| U |\circlearrowright \rangle = \frac{1}{\sqrt{2}} \left ( y \rme^{\rmi \varphi}\cos \frac{\theta}{2} \sigma_0 + \sin \frac{\theta}{2} \sigma_Y \right ) ,
\end{equation}
and  
\begin{equation}
    \Lambda^\dagger_y \Lambda_y = \Delta_Y (y) = \frac{1}{2} \left ( \sigma_0+ y \gamma_Y \sigma_Y \right ) .
\end{equation}
Then it holds 
\begin{equation}
    p(x|y) = \mathrm{tr}_S \left [\tilde{\rho}_y |x \rangle \langle x| \right ] ,
\end{equation}
which is different from the relation we were looking for
\begin{equation}
    p(x|y) = \mathrm{tr}_S \left [\rho_y \Delta_X (x) \right ].
\end{equation}

\bigskip

\section{Number states in double homodyne detection}

Let us present an example beyond the qubit system that may illustrate the application of our approach in system spaces with higher dimension. This can be the case of double homodyne detection in an electromagnetic field mode of complex-amplitude operator $a$. The double homodyne detection may be considered as a joint simultaneous measurement of two field quadratures $X$ and $Y$ \cite{WC86,NGW87}
\begin{equation}
X = \frac{1}{2} \left ( a + a^\dagger \right ) , \qquad Y = \frac{i}{2} \left ( a^\dagger - a \right ),
\end{equation}
whose statistics is given essentially, up to dilation factors, by the Husimi Q function \cite{MW95}. So, the POVM in this case is 
\begin{equation}
    \Delta (x,y) = \frac{1}{\pi} |\alpha \rangle \langle \alpha |, \qquad \alpha = x + i y, 
\end{equation}
where $ |\alpha \rangle$ are the Glauber-Sudarshan coherent states, and here $x$, $y$ are unbounded continuous Cartesian variables. Let the mode be in a one-photon state, so that 
\begin{equation}
    p(x,y) = \frac{1}{\pi} (x^2 + y^2)e^{- x^2- y^2} , \quad  p_Y (y) = \frac{1}{2 \sqrt{\pi}} (1+ 2y^2)e^{- y^2} ,
\end{equation}
and the conditioned probability is 
\begin{equation}
    p(x|y) = \frac{2}{\sqrt{\pi}} \frac{x^2 + y^2}{1+2y^2}e^{- x^2} .
\end{equation}
Let us show that we can find values of $y$ such that $\rho_y \nsucceq 0$. To this end we note that 
\begin{equation}
    \Delta_X (x) = \sqrt{\frac{2}{\pi}} \int_{-\infty}^\infty dx^\prime e^{-2(x-x^\prime)^2} |x^\prime \rangle \langle x^\prime | ,
\end{equation}
where $|x \rangle$ are the eigenstates of the quadrature $X$, 
so that for any $\rho_y$
\begin{equation}
\mathrm{tr} \left [\rho_y \Delta_X (x) \right ] =  \sqrt{\frac{2}{\pi}} \int_{-\infty}^\infty dx^\prime e^{-2(x-x^\prime)^2}p_y (x^\prime) ,
\end{equation}
where $p_y (x) = \langle x |\rho_y |x \rangle $ is the probability distribution of the $X$ quadrature in the $\rho_y$ state. The question is whether the equality  $p(x|y) = \mathrm{tr} \left [\rho_y \Delta_X (x) \right ]$ can be fulfilled with a legitimate $p_y (x)$ distribution. The solution is simple since the equality can be inverted via Fourier analysis to determine $p_y (x)$ with the result:
\begin{equation}
    p_y(x) = \sqrt{\frac{2}{\pi}} \frac{8 x^2+2 y^2-1}{1+2y^2}e^{- 2 x^2}.
\end{equation}
We can see that whenever $y^2 < 1/2$ the distribution $p_y(x) = \langle x |\rho_y |x \rangle$ takes negative values, for example at $x=0$, and then $\rho_y \nsucceq 0$.

\end{document}